\documentclass{article}
\usepackage{amsmath,amssymb,a4wide}
\advance\topmargin -1.5cm

\newcommand{\FA}{\textsl{FeynArts}}
\newcommand{\FC}{\textsl{FormCalc}}
\newcommand{\FO}{\textsl{FORM}}
\newcommand{\LT}{\textsl{LoopTools}}
\newcommand{\mma}{\textsl{Mathematica}}
\newcommand{\lbrac}{\symbol{123}}
\newcommand{\rbrac}{\symbol{125}}
\newcommand{\uscore}{\symbol{95}}
\newcommand{\eg}{e.g.\ }
\newcommand{\ie}{i.e.\ }
\newcommand{\grey}[1]{\special{ps:.7 setgray}#1\special{ps:0 setgray}}

\renewcommand{\O}{\mathcal{O}}
\newcommand{\sw}{s_W}
\newcommand{\cw}{c_W}
\newcommand{\mtbar}{\overline{m}_t}
\newcommand{\MSUSY}{M_{\text{SUSY}}}
\newcommand{\MSbar}{\overline{\text{MS}}}

\newcommand{\GeV}{\text{ GeV}}
\newcommand{\unity}{{\rm 1\mskip-4.25mu l}}
\newcommand{\mssmini}{{\tt mssm\uscore ini.F}}

\newcommand{\sinb}{\sin\beta}
\newcommand{\cosb}{\cos\beta}
\newcommand{\tanb}{\tan\beta}
\newcommand{\cotb}{\cot\beta}
\newcommand{\sinQb}{\sin^2\!\beta}
\newcommand{\cosQb}{\cos^2\!\beta}

\DeclareMathOperator{\diag}{diag}
\DeclareMathOperator{\Li}{Li}

\renewcommand{\arraystretch}{1.2}

\allowdisplaybreaks
\makeatletter

\def\reportno#1{\gdef\@reportno{#1}}

\def\@maketitle{%
  \hfill{\small\begin{tabular}[t]{r}%
    \@reportno
  \end{tabular}\par}%
  \vskip 2em%
  \begin{center}%
    \let\footnote\thanks%
    {\large\@title\par}%
    \vskip 1.5em%
    \lineskip .5em%
    \begin{tabular}[t]{c}%
      \@author
    \end{tabular}\par%
    \vskip 1em%
    \@date%
  \end{center}%
  \par
  \vskip 1.5em}

\makeatother

\begin{document}

\title{The Implementation of the Minimal Supersymmetric Standard Model\\
	in \FA\ and \FC}

\author{Thomas Hahn, Christian Schappacher \\
	{\it Institut f\"ur Theoretische Physik} \\
	{\it Universit\"at Karlsruhe} \\
	{\it D--76128 Karlsruhe, Germany}}

\reportno{KA--TP--18--2001\\hep-ph/0105349}

\date{June 1, 2001}

\maketitle

\begin{abstract}
We describe the implementation of the MSSM in the diagram generator \FA\
and the calculational tool \FC. This extension allows to perform loop
calculations of MSSM processes almost fully automatically. The actual
implementation has two aspects: The MSSM Feynman rules are specified in a
new model file for \FA. The computation of the parameters in the MSSM
Lagrangian from the input parameters is realized as a Fortran subroutine
in the framework of \FC. The model file does not depend on the latter,
however, and can be used even if one does not want to continue the
calculation with \FC. The Feynman rules have been entered in a very
generic way to allow \eg scenarios with complex parameters, and have been
tested extensively by reproducing known results for several non-trivial
scattering processes. \\[1ex]
\end{abstract}


\section{Introduction}

One of the main problems of Feynman-diagrammatic computations is the
enormous growth of the number of Feynman diagrams, not only with the loop
order, but also with the number of particles in a model. While many
precision calculations in the Standard Model (SM) could still be performed
by hand exactly, the same is very difficult in models like the Minimal
Supersymmetric Standard Model (MSSM). Yet it is highly desirable to
perform unabridged calculations in the MSSM, too, since also the MSSM
allows to make precise predictions in terms of a set of input parameters.

With the availability of powerful software packages, the basic problem of
bookkeeping and calculation of the diagrams has been solved for many
common cases. Still, it is not entirely trivial to code a model of the
complexity of the MSSM in such a system, since this has to be done in a
reasonably general way (\ie not only for special cases of the parameters)
and many checks have to be performed to test all sectors of the model.

The present paper documents the implementation of the MSSM in the \FA\
\cite{KuBD90} and \FC\ \cite{HaP98} packages. Other programs for which an
MSSM model file exists are GRACE \cite{Yu00, Ku99} and CompHEP
\cite{BoDIPS94, BeGS97}. Conceptually, the \FA--\FC\ system works in three
stages, as sketched in the following diagram.
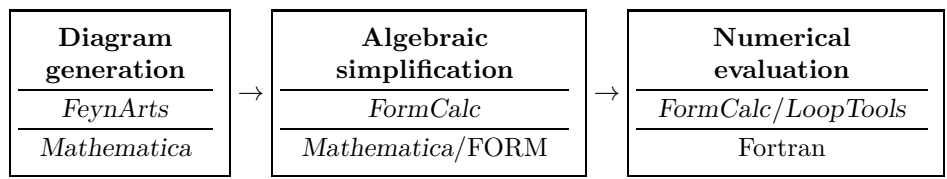
\begin{equation*}
\fbox{%
\begin{tabular}{c}
{\bf Diagram} \\[-.5ex]
{\bf generation} \\ \hline
\FA \\ \hline
\mma
\end{tabular}
}\to\fbox{%
\begin{tabular}{c}
{\bf Algebraic} \\[-.5ex]
{\bf simplification} \\ \hline
\FC \\ \hline
\mma/FORM
\end{tabular}
}\to\fbox{%
\begin{tabular}{c}
{\bf Numerical} \\[-.5ex]
{\bf evaluation} \\ \hline
\FC/\LT \\ \hline
Fortran
\end{tabular}
}
\end{equation*}
A more detailed discussion of the interplay between \FA\ and \FC\ can be
found in \cite{Ha00}.

This paper is organized as follows. Corresponding to the three stages,
Diagram generation, Algebraic simplification, and Numerical evaluation,
there are three sections which describe the modifications to each stage
that are necessary for calculations in the MSSM.

Sect.\ \ref{sect:diaggen} describes the changes in \FA, namely the new
MSSM model file. The model file declares the properties of the fields,
their propagators, and their couplings. It contains the parameters of the
MSSM Lagrangian. The Feynman rules have been entered in a very generic way
to allow \eg scenarios with complex parameters.

Sects.\ \ref{sect:algsimp} and \ref{sect:numeval} describe the changes in
\FC. Sect.\ \ref{sect:algsimp} outlines the changes in the algebraic
simplification and Sect.\ \ref{sect:numeval} explains the calculation of
the parameters that are used in the model file. The model-file parameters
are derived from a (reasonably small) set of input parameters. This is
realized as a model-initializing Fortran subroutine in the framework of
\FC. The model file does not depend on the latter, however, and can be
used even if one does not want to continue the calculation with \FC.


\section{Diagram generation}
\label{sect:diaggen}

\subsection{Algorithms}

Supersymmetric theories contain fermion-number-violating couplings, \eg
quark--squark--gluino, so that the usual method of ordering the Dirac
matrices oppositely to their occurrence along the arrows on fermionic
lines breaks down since one cannot define a fermion-number flow.

\FA\ uses the ``flipping-rule'' algorithm \cite{DeEHK92}. This algorithm
was invented precisely to solve the problem of fermion-number-violating
couplings and works as follows: Instead of traversing the fermion lines
along the fermion-number flow imposed from the outside, \FA\ chooses a
direction for each fermion chain. If it turns out later that, for a Dirac
fermion, the chosen direction is opposite to the actual fermion flow, \FA\
``flips'' the coupling, \ie it derives the coupling appropriate for the
reversed fermion flow from the known coupling. The flipping of a coupling
is in fact nothing but a charge (as opposed to hermitian) conjugation.


\subsection{The MSSM model file}

The model file is the source of all physics information in \FA. It
declares the properties of the fields, their propagators, and their
couplings. In the model file the parameters of the Lagrangian are used,
not a restricted set of input parameters.

There are two versions of the MSSM model file in \FA, both of which follow
the conventions of \cite{HaK85, GuH86, hhg}. The file {\tt MSSMQCD.mod}
defines the complete (electroweak and strong) MSSM, whereas {\tt MSSM.mod}
contains only the electroweak subset, defined as everything except the
gluon, its ghost, and the gluino. The four-sfermion couplings appear in
{\tt MSSM.mod} although they have both electroweak and strong parts.
Counter-terms are not yet entered in the model files.

Table \ref{tab:mssmparticles} gives the names for the fields and
corresponding masses defined in {\tt MSSM.mod} and {\tt MSSMQCD.mod}. The
symbols used for the MSSM parameters are specified in Table
\ref{tab:modelsyms}. The complete list of couplings is too long to be
included here, but is contained in the \FA\ distribution as a PostScript
file.

The MSSM model files further declare a number of restrictions with which
certain groups of particles can be excluded in the diagram generation.  
These restrictions are listed in Table \ref{tab:restrict}.

\begin{table}
\begin{center}
\begin{tabular}{|l|l|l|l||l|l|l|l|} \hline
leptons & $f = f^\dagger$ & field & mass &
sleptons & $f = f^\dagger$ & field & mass \\ \hline
$\nu_g$ & & {\tt F[1,\,\lbrac$g$\rbrac]} & {\tt 0} &
  $\tilde\nu_g$ & & {\tt S[11,\,\lbrac$g$\rbrac]} & {\tt MSf} \\
$\ell_g$ & & {\tt F[2,\,\lbrac$g$\rbrac]} & {\tt MLE} &
  $\tilde\ell_g^s$ & & {\tt S[12,\,\lbrac$s,g$\rbrac]} & {\tt MSf} \\
\hline
\multicolumn{6}{c}{} \\[-1ex] \hline
\multicolumn{4}{|l||}{quarks} &
\multicolumn{4}{|l|}{squarks} \\ \hline
$u_g$ & & {\tt F[3,\,\lbrac$g,o$\rbrac]} & {\tt MQU} &
  $\tilde u_g^s$ & & {\tt S[13,\,\lbrac$s,g,o$\rbrac]} & {\tt MSf} \\
$d_g$ & & {\tt F[4,\,\lbrac$g,o$\rbrac]} & {\tt MQD} &
  $\tilde d_g^s$ & & {\tt S[14,\,\lbrac$s,g,o$\rbrac]} & {\tt MSf} \\
\hline
\multicolumn{6}{c}{} \\[-1ex] \hline
\multicolumn{4}{|l||}{gauge bosons} &
\multicolumn{4}{|l|}{neutralinos, charginos} \\ \hline
$\gamma$ & yes & {\tt V[1]} & {\tt 0} &
  $\tilde\chi_n^0$ & yes & {\tt F[11,\,\lbrac$n$\rbrac]} & {\tt MNeu} \\
$Z$ & yes & {\tt V[2]} & {\tt MZ} &
  $\tilde\chi_c^-$ & & {\tt F[12,\,\lbrac$c$\rbrac]} & {\tt MCha} \\ 
$W^-$ & & {\tt V[3]} & {\tt MW} & & & & \\ \hline
\multicolumn{6}{c}{} \\[-1ex] \hline
\multicolumn{4}{|l||}{Higgs bosons} &
\multicolumn{4}{|l|}{ghosts} \\ \hline
$h^0$ & yes & {\tt S[1]} & {\tt Mh0} &
	$u_\gamma$ & & {\tt U[1]} & {\tt 0} \\
$H^0$ & yes & {\tt S[2]} & {\tt MHH} &
	$u_Z$ & & {\tt U[2]} & {\tt MZ} \\
$A^0$ & yes & {\tt S[3]} & {\tt MA0} &
	$u_+$ & & {\tt U[3]} & {\tt MW} \\
$G^0$ & yes & {\tt S[4]} & {\tt MG0} &
	$u_-$ & & {\tt U[4]} & {\tt MW} \\
$H^-$ & & {\tt S[5]} & {\tt MHp} &
	\grey{$u_g$} & & \grey{\tt U[5,\,\lbrac$u$\rbrac]} & \grey{\tt 0} \\
$G^-$ & & {\tt S[6]} & {\tt MGp} & & & & \\ \hline
\multicolumn{6}{c}{} \\[-1ex] \hline
\multicolumn{4}{|l||}{\grey{gluon}} &
\multicolumn{4}{|l|}{\grey{gluino}} \\ \hline
\grey{$g$} & \grey{yes} & \grey{\tt V[5,\,\lbrac$u$\rbrac]} &
  \grey{\tt 0} &
  \grey{$\tilde g$} & \grey{yes} & \grey{\tt F[15,\,\lbrac$u$\rbrac]} &
  \grey{\tt MGl} \\
\hline
\end{tabular}
\end{center}
where the following indices are used:
\begin{alignat*}{4}
g &= \text{\tt Index[Generation]} &&= 1\dots 3\,, & \qquad\qquad
s &= \text{\tt Index[Sfermion]}   &&= 1\dots 2\,, \\
o &= \text{\tt Index[Colour]}     &&= 1\dots 3\,, &
n &= \text{\tt Index[Neutralino]} &&= 1\dots 4\,, \\
u &= \text{\tt Index[Gluon]}      &&= 1\dots 8\,, &
c &= \text{\tt Index[Chargino]}   &&= 1\dots 2\,.
\end{alignat*}
\caption{\label{tab:mssmparticles}The particle set-up in {\tt MSSM.mod}
and {\tt MSSMQCD.mod}. The gluon, its ghost, and the gluino, which are
defined only in the latter, are written in grey.}
\end{table}

\begin{table}
\begin{center}
\begin{tabular}{|l|l|} \hline
{\tt Mh0, MHH, MA0, MG0} &
	neutral Higgs masses \\
{\tt MHp, MGp} &
	charged Higgs masses \\
{\tt CB, SB, TB} &
	$\cos\beta$, $\sin\beta$, $\tan\beta$ \\
{\tt CA, SA} &
	$\cos\alpha$, $\sin\alpha$ \\
{\tt C2A, S2A, C2B, S2B} &
	$\cos 2\alpha$, $\sin 2\alpha$,
	$\cos 2\beta$, $\sin 2\beta$ \\
{\tt CAB, SAB, CBA, SBA} &
	$\cos(\alpha + \beta)$, $\sin(\alpha + \beta)$,
	$\cos(\beta - \alpha)$, $\sin(\beta - \alpha)$ \\
{\tt MUE} &
	Higgs-doublet mixing parameter $\mu$ \\ \hline
{\tt MGl} &
	gluino mass \\
{\tt MNeu[$n$]} &
	neutralino masses \\
{\tt ZNeu[$n$,\,$n'$]} &
	neutralino mixing matrix \\
{\tt MCha[$c$]} &
	chargino masses \\
{\tt UCha[$c$,\,$c'$], VCha[$c$,\,$c'$]} &
	chargino mixing matrices \\ \hline
{\tt MSf[$s$,\,$t$,\,$g$]} &
	sfermion masses \\
{\tt USf[$t$,\,$g$][$s$,\,$s'$]} &
	sfermion mixing matrix \\
{\tt Af[$t$,\,$g$] } &
	(scalar) soft-breaking $A$-parameters \\ \hline
\end{tabular}
\end{center}
The indices enumerate the following properties:
\begin{equation*}
\begin{aligned}[t]
& \text{sfermion generation:} \quad g = 1\dots 3\,, \\
& \text{sfermion type:} \quad t = \begin{cases}
  1 & \text{sneutrinos}\,, \\
  2 & \text{sleptons}\,, \\
  3 & \text{up-type squarks}\,, \\
  4 & \text{down-type squarks}\,,
\end{cases}
\end{aligned}
\qquad
\begin{aligned}[t]
\text{eigenstates:} &&& \text{mass} && \text{gauge} \\
\text{sfermion} &\quad &
	s &= 1\dots 2\,, &\quad
	s' &= 1\dots 2\,, \\
\text{neutralino} & &
	n &= 1\dots 4\,, &
	n' &= 1\dots 4\,, \\
\text{chargino} & &
	c &= 1\dots 2\,, &
	c' &= 1\dots 2\,. \\
\end{aligned}
\end{equation*}
\caption{\label{tab:modelsyms}Symbols representing the MSSM parameters
introduced by {\tt MSSM.mod} and {\tt MSSMQCD.mod}.}
\end{table}

\begin{table}
\begin{center}
\begin{tabular}{|l|p{.6\linewidth}|} \hline
{\tt NoGeneration1} &
	exclude generation-1 fermions
	($\nu_{\rm e}$, $e$, $u$, $d$) \\
{\tt NoGeneration2} &
	exclude generation-2 fermions
	($\nu_\mu$, $\mu$, $c$, $s$) \\
{\tt NoGeneration3} &
	exclude generation-3 fermions
	($\nu_\tau$, $\tau$, $t$, $b$) \\ \hline
{\tt NoElectronHCoupling} &
	exclude all couplings involving electrons and any Higgs
	field \\
{\tt NoLightFHCoupling} &
	exclude all couplings between light fermions
	(all fermions except the top) and any Higgs field \\ \hline
{\tt NoSUSYParticles} &
	exclude the particles not present in the SM: sfermions,
	charginos, neutralinos, and the Higgs fields
	$H^0$, $A^0$, $H^\pm$ \\
{\tt THDMParticles} &
	exclude the particles not present in the two-Higgs-doublet
	model: the sfermions, charginos, and neutralinos \\ \hline
\end{tabular}
\end{center}
\caption{\label{tab:restrict}Pre-defined restrictions in {\tt MSSM.mod}
and {\tt MSSMQCD.mod}.}
\end{table}


\section{Algebraic simplification}
\label{sect:algsimp}


\subsection{Algorithms}

The algebraic simplification has to address two problems that arise in
one-loop calculations in supersymmetric theories in general and in the
MSSM in particular. The first is the conceptual problem of implementing a
supersymmetry-preserving regularization scheme, the second is the
technical problem of dealing with more diagrams and larger expressions.

The default regularization scheme employed by \FC\ is dimensional
regularization. Unfortunately, this scheme is known to break supersymmetry
\cite{CaJN80}. \FC\ has therefore been equipped with an alternative scheme
suited for calculations in supersymmetric theories. It is the constrained
differential renormalization scheme (CDR) \cite{AgCMP99} which is
equivalent to regularization by dimensional reduction at the one-loop
level \cite{HaP98}. The schemes are chosen with the {\tt Dimension} option
of the {\tt CalcFeynAmp} function: {\tt Dimension -> D} selects
dimensional regularization (the default), {\tt Dimension -> 4} switches to
CDR.

The second problem is a technical one. In the SM, both the number of
diagrams and the comparatively simple coupling structure allow to perform
calculations with ``brute force.'' To wit, the diagrams are generated with
all indices explicitly written out, and then calculated.

The situation is less favourable in the MSSM, however: not only is the
number of diagrams considerably higher in most cases, but also the
coupling structure is more involved. This is because particles like
sfermions, or gauginos and higgsinos in general mix to form mass
eigenstates and hence their couplings contain plenty of mixing matrices
and can become rather lengthy. This means that the algebraic
simplification has to be performed more carefully in order to maintain a
decent performance and keep the size of the results as small as possible.

The way in which \FC\ proceeds is already suggested by the level structure
of the \FA\ amplitudes. The lowest (generic) level completely determines
the kinematical structure of a diagram, so \FC\ first performs the
time-consuming simplifications involving kinematical quantities, like the
tensor reduction, on the far fewer generic diagrams only. For each diagram
at the next higher (classes) level, it then substitutes the generic
coupling constants and masses by their actual values. The diagrams are
mostly specified by then, except that index summations, \eg over fermion
generations, are not yet carried out. These index summations are performed
only in the Fortran code, which means that \FC\ has to keep track of all
indices and write out the corresponding loop instructions in Fortran. For
a model like the MSSM, the savings incurred with this method of
simplification may easily amount to an order of magnitude in both CPU time
and size of the Fortran code.


\section{Numerical evaluation}
\label{sect:numeval}

In the \FC\ framework, the \mma\ expressions resulting from the algebraic
simplification are translated to Fortran code for numerical evaluation.
The generated Fortran code of course has to be provided with the proper
numerical values for the parameters appearing in the model, \ie the
variables in Table \ref{tab:modelsyms}. This is solved by a subroutine
which is called at the beginning of the calculation to initialize all
model parameters.

Technically, the input parameters are specified in common blocks defined
in {\tt model.h}. Some less commonly changed inputs, notably some breaking
parameters in the sfermion sector, are realized as preprocessor variables
which take default values if not defined by the user. Preprocessor
variables are also used for switches, such as whether squark mixing should
be turned on or off.

\subsection{Parameters of the MSSM}

Supersymmetry (SUSY) completely determines the supersymmetric part of the
MSSM Lagrangian once the SM parameters are known. The complete MSSM,
however, with softly broken SUSY in its general form, necessarily
introduces a large number of masses, phases, and mixing angles to
parametrize the SUSY breaking. At a final count, 105 of these degrees of
freedom cannot be absorbed in some other quantity or rotated away
\cite{DiS95}. In order to get a handle on so many parameters, several
assumptions have been made in \mssmini\ to arrive at a moderate number of
input parameters while retaining reasonable generality.

The first assumption is that the SUSY-GUT relation holds, which relates
the U(1), SU(2), and SU(3) gaugino mass parameters $M_1$, $M_2$, and $M_3$
according to
\begin{equation}
M_1 = \frac 53\frac{\sw^2}{\cw^2} M_2
\quad\text{and}\quad
M_3 = \frac{\overline{\alpha}_s(s)}{\overline{\alpha}(s)}\sw^2 M_2
\quad\text{with}\quad
m_{\tilde g}\equiv |M_3|\,,
\end{equation}
where $\sqrt s$ is the CMS energy, $\cw = M_W/M_Z$, and
$\sw = \sqrt{1 - \cw^2}$. The running couplings $\overline{\alpha}$ and 
$\overline{\alpha}_s$ are $\MSbar$ values; we use the CERNlib function
{\tt ALPHAS2} \cite{pdflib} for $\overline{\alpha}_s(s)$ and
\begin{equation}
\frac 1{\overline{\alpha}(s)} =
  \frac 1{\overline{\alpha}(M_Z^2)} -
  \frac 1{3\pi}\sum_{f\neq t} Q_f^2 N_c\ln\frac s{M_Z^2}
= \frac 1{\overline{\alpha}(M_Z^2)} -
  \frac{20}{9\pi}\ln\frac s{M_Z^2}
\end{equation}
with $1/\overline{\alpha}(M_Z^2) = 127.934$ \cite{PDG}.

Secondly, we assume the SUSY-breaking parameters in the sfermion sector to
be flavour-blind and distinguish only two $A$-parameters for different
isospin values, hence the sfermion sector is governed by seven scalar
parameters:
\begin{equation}
\begin{aligned}
{\bf M}^2_{\tilde Q} &= M^2_{\tilde Q}\unity\,, &\qquad
{\bf M}^2_{\tilde U} &= M^2_{\tilde U}\unity\,, &\qquad
{\bf A}_U &= A_u\unity\,, \\
& &
{\bf M}^2_{\tilde D} &= M^2_{\tilde D}\unity\,, &
{\bf A}_D &= A_d\unity\,, \\
{\bf M}^2_{\tilde L} &= M^2_{\tilde L}\unity\,, &
{\bf M}^2_{\tilde E} &= M^2_{\tilde E}\unity\,, &
{\bf A}_L &= A_d\unity\,,
\end{aligned}
\end{equation}
where in addition the various $M_{\tilde X}$ get the default value
$\MSUSY$, a common SUSY-breaking mass, \ie
\begin{equation}
M_{\tilde Q} = M_{\tilde L}
= M_{\tilde U} = M_{\tilde D} = M_{\tilde E} = \MSUSY
\quad\text{(by default)}\,.
\end{equation}
The remaining MSSM input parameters thus have 10 degrees of freedom:
\begin{equation}
\begin{array}{lll}
1 & \text{quotient of {\sc vev}'s:} & \tanb\ \text{(real)}, \\
1 & \text{Higgs mass:} & M_A\ \text{(real)}, \\
6 & \text{breaking parameters:}	&
	A_u, A_d\ \text{(complex)},\ \MSUSY, M_2\ \text{(real)}, \\
2 & \text{mixing parameter:} & \mu\ \text{(complex)}.
\end{array}
\end{equation}
The inputs to \mssmini\ are summarized in Table \ref{tab:mssmini}. All
other variables in the MSSM Lagrangian (see Table \ref{tab:modelsyms})
are determined once this reduced set of MSSM input parameters and the SM
inputs are specified and are calculated by \mssmini. This is the reason
why \mssmini\ is much more involved than its companion file for the SM,
{\tt sm\uscore ini.F}. The calculation of the parameters can be directed
by several switches, which are also listed in Table \ref{tab:mssmini}.

\begin{table}
\begin{center}
\begin{tabular}{|l|l|l|l|} \hline
parameter & Fortran name & type & default value \\ \hline
$\tanb$ & {\tt TB} & double precision & \\
$M_A$ & {\tt MA0} & double precision & \\
$A_u$ & {\tt Au} & double complex & \\
$A_d$ & {\tt Ad} & double complex & \\
$\MSUSY$ & {\tt MSusy} & double precision & \\
$M_2$ & {\tt M\uscore 2} & double precision & \\
$\mu$ & {\tt MUE} & double complex & \\
$M_{\tilde Q}$ & {\tt MSQ} & preprocessor variable & {\tt MSusy} \\
$M_{\tilde L}$ & {\tt MSL} & preprocessor variable & {\tt MSusy} \\
$M_{\tilde U}$ & {\tt MSU} & preprocessor variable & {\tt MSusy} \\
$M_{\tilde D}$ & {\tt MSD} & preprocessor variable & {\tt MSusy} \\
$M_{\tilde E}$ & {\tt MSE} & preprocessor variable & {\tt MSusy} \\ \hline
\end{tabular}

\bigskip

\begin{tabular}{|l|p{.67\linewidth}|} \hline
switch & action \\ \hline
{\tt NO\uscore SQUARK\uscore MIXING} &
	Sets $A_u = \mu^*\cotb$ and $A_d = \mu^*\tanb$, so that the 
	off-diagonal entries of the sfermion mass matrices vanish (cf.\
	Eq.\ \eqref{eq:nosqmix}), \ie makes sfermion mass eigenstates =
	sfermion gauge eigenstates. \\ \hline
{\tt COMPLEX\uscore PARAMETERS} &
	Uses a simpler (one-loop) approximation for the Higgs masses
	which is valid for all parameters. This switch must be set if
	complex input parameters are used because otherwise the more
	precise (two-loop) approximation for the Higgs masses is taken,
	which is valid only for real parameters. \\ \hline
{\tt SM\uscore ONLY} &
	Calculates $M_h$ as usual, but then reverse-engineers the mixing
	in the Higgs sector ($\alpha$ and $\beta$) such that the MSSM
	Higgs sector looks like a SM Higgs sector (see \cite{hhg}, p.\
	356), with the light CP-even Higgs boson $h$ figuring as the SM
	Higgs boson. \\ \hline
{\tt NO\uscore EXCLUSION\uscore LIMITS} &
	Ignores the experimental bounds, \ie does not exclude points in
	parameter space if the bounds in Eq.\ \eqref{eq:bounds} are
	violated. \\ \hline
\end{tabular}
\end{center}
\caption{\label{tab:mssmini}The inputs and switches for \mssmini.
Note: A preprocessor variable is defined with ``{\tt\#define} {\it var}
{\it value}'' and a switch is set with ``{\tt\#define} {\it switch}''.
The {\tt\#define} must stand at the beginning of the line.}
\end{table}

When the particle masses are calculated from these input parameters,
\mssmini\ checks whether they are consistent with the current exclusion
limits and automatically omits already-excluded points with a warning. The
following bounds are used:
\begin{equation}
\label{eq:bounds}
\begin{aligned}
m_{\tilde t} &\geqslant 80\GeV~\cite{squarkbounds}\,, \\
m_{\tilde b} &\geqslant 70\GeV~\cite{squarkbounds}\,, \\
m_{\tilde q\neq\tilde b,\tilde t} &\geqslant 150\GeV~\cite{squarkbounds}\,, \\
m_{\tilde\ell} &\geqslant 70\GeV~\cite{sleptonbounds}\,, \\
\Delta\rho_{\tilde t, \tilde b} &\leqslant 3\times 10^{-3}~\cite{PDG}\,,
\end{aligned}
\qquad
\begin{aligned}
M_h &\geqslant
\begin{cases}
  91\GeV~\text{for real input parameters}~\cite{h0bounds}\,, \\
  85\GeV~\text{for complex input parameters}~\cite{h0boundscplx}\,, \\
\end{cases} \\
m_{\tilde\chi} &\geqslant 90\GeV~\cite{chibounds}\,, \\
m_{\tilde\chi^0} &\geqslant 30\GeV~\cite{chibounds}\,, \\
m_{\tilde g} &\geqslant 175\GeV~\cite{gluinobounds}\,.
\end{aligned}
\end{equation}

The detailed formulas built into \mssmini\ for computing the
various parameters are presented in the following sections, organized
according to the various sectors of the MSSM.


\subsection{The Higgs sector}

The neutral Higgs sector is fixed by choosing a value for $\tanb =
v_2/v_1$ (the ratio of the vacuum expectation values of the two Higgs
doublets) and for the mass $M_A$ of the CP-odd neutral Higgs boson $A^0$.
For the CP-even Higgs masses, which receive sizable radiative corrections,
we use the approximation formula of \cite{HeHW99} which agrees with the
full two-loop calculation \cite{HeHW99b} to within less than 2 GeV.

\subsubsection{The neutral CP-even Higgs bosons $h$ and $H$}

In this section we discuss the case of real MSSM parameters, for which a
more sophisticated computation of the Higgs masses is implemented. The
case of complex parameters is treated in Sect.\ \ref{sect:complexpara}.

The gauge eigenstates $\phi_1$ and $\phi_2$ of the neutral CP-even Higgs
bosons mix via the mass matrix
\begin{equation}
M^2_{\text{Higgs}} = \begin{pmatrix}
M_A^2\sinQb + M_Z^2\cosQb - \hat\Sigma_{\phi_1} &
	-(M_A^2 + M_Z^2)\sinb\cosb - \hat\Sigma_{\phi_1\phi_2} \\
-(M_A^2 + M_Z^2)\sinb\cosb - \hat\Sigma_{\phi_1\phi_2} &
	M_A^2\cosQb + M_Z^2\sinQb - \hat\Sigma_{\phi_2}
\end{pmatrix}.
\end{equation}
Diagonalizing this matrix yields the Higgs masses
\begin{align}
M_{H,h}^2 &=
\frac{M_A^2 + M_Z^2 - \hat\Sigma_{\phi_1} - \hat\Sigma_{\phi_2}}{2}
\pm \biggl[
  \frac{(M_A^2 + M_Z^2)^2 +
    (\hat\Sigma_{\phi_1} - \hat\Sigma_{\phi_2})^2}{4}
  - M_A^2 M_Z^2 \cos^2 2\beta
  \nonumber \\
&\qquad\quad
  + \frac 12 (M_A^2 - M_Z^2)\cos 2\beta\,
      (\hat\Sigma_{\phi_1} - \hat\Sigma_{\phi_2})
  + (M_A^2 + M_Z^2)\sin 2\beta\,\hat\Sigma_{\phi_1\phi_2}
  + \hat\Sigma_{\phi_1\phi_2}^2
\biggr]^{1/2}.
\end{align}
The mixing angle $\alpha$ of the CP-even Higgs doublet is hence
\begin{equation}
\alpha = \arctan\!\left(
  \frac{-(M_A^2 + M_Z^2)\sinb\cosb - \hat\Sigma_{\phi_1\phi_2}}
       {M_Z^2\cosQb + M_A^2\sinQb - \hat\Sigma_{\phi_1} - M_h^2}
\right).
\end{equation}
The presence of renormalized self-energies $\hat\Sigma$ indicates that the
mass matrix contains significant radiative corrections which must be taken
into account to make quantitatively correct predictions. The self-energies
of course receive contributions from all sectors of the MSSM, but not all
are numerically of equal relevance. We take into account only the
following terms:
\begin{itemize}
\item
$\hat\Sigma^{(1, t/\tilde t)}(0)$: the one-loop $t/\tilde t$-contributions
at zero momentum transfer up to $m_t^4$ \cite{Da95},

\item
$\hat\Sigma^{(2, t/\tilde t)}(0)$: the dominant two-loop
$t/\tilde t$-contributions of $\O(\alpha\alpha_s)$ at zero momentum 
transfer \cite{HeHW99} and the leading two-loop Yukawa correction
of $\O(\alpha^2)$ \cite{CaEQW95},

\item
$\hat\Sigma^{(1, \text{rest})}$: the one-loop leading-log contributions
from all other sectors \cite{HaHH97}.
\end{itemize}
In the following, we give the explicit expressions for these contributions
as implemented in \mssmini.
\begin{itemize}
\item
The one-loop $t/\tilde t$-contributions at zero momentum transfer up
to $m_t^4$:
\begin{align}
\hat\Sigma^{(1, t/\tilde t)}_{\phi_1}(0) &=
  \frac{G_F\sqrt 2}{\pi^2} M_Z^4\Lambda\cosQb
  \ln\frac{m_t^2}{M_S^2}\,, \\
\hat\Sigma^{(1, t/\tilde t)}_{\phi_1\phi_2}(0) &=
  -\frac{G_F\sqrt 2}{\pi^2} M_Z^2\cotb 
  \left(-\frac 38 m_t^2 + M_Z^2\Lambda\sinQb\right)
  \ln\frac{m_t^2}{M_S^2}\,, \\
\label{eq:sigphi2}
\hat\Sigma^{(1, t/\tilde t)}_{\phi_2}(0) &=
\frac{G_F\sqrt 2}{\pi^2}\frac{m_t^4}{8\sinQb} \Biggl[
 -2\frac{M_Z^2}{m_t^2} + \frac{11}{10}\frac{M_Z^4}{m_t^4}
  \nonumber \\
&\qquad
  + \left(12 - 6\frac{M_Z^2}{m_t^2}\sinQb
       + 8\frac{M_Z^4}{m_t^4}\Lambda\sin^4\!\beta\right)
  \ln\frac{m_t^2}{M_S^2}
  \nonumber \\
&\qquad
  + \left(-12 + 4\frac{M_Z^2}{m_t^2} + 6\frac{m_t^2}{M_S^2}\right)
    \frac{(M_t^{LR})^2}{M_S^2}
  + \left(1 - 4\frac{m_t^2}{M_S^2} + 3\frac{m_t^4}{M_S^4}\right)
    \frac{(M_t^{LR})^4}{M_S^4}
  \nonumber \\
&\qquad
  + \left(\frac 35\frac{m_t^2}{M_S^2}
      - \frac{12}{5}\frac{m_t^4}{M_S^4} + 2\frac{m_t^6}{M_S^6}\right)
    \frac{(M_t^{LR})^6}{M_S^6}
  \nonumber \\
&\qquad
  + \left(\frac 37\frac{m_t^4}{M_S^4}
      - \frac{12}{7}\frac{m_t^6}{M_S^6}
      + \frac 32\frac{m_t^8}{M_S^8} \right)
    \frac{(M_t^{LR})^8}{M_S^8}
  \Biggr]
\end{align}
with
\begin{align}
M_t^{LR} &= A_u - \mu^*\tanb\,, \\
M_S &= \left[M_{\tilde Q}^2 M_{\tilde U}^2
  + m_t^2 (M_{\tilde Q}^2 + M_{\tilde U}^2) + m_t^4\right]^{1/4}, \\
\Lambda &= \frac 18 - \frac 13\sw^2 + \frac 49\sw^4\,.
\end{align}

\item
The dominant two-loop $t/\tilde t$-contributions of $\O(\alpha\alpha_s)$
at zero momentum transfer and the leading two-loop Yukawa correction
of $\O(\alpha^2)$:
\begin{align}
\hat\Sigma^{(2, t/\tilde t)}_{\phi_1}(0) &=
  \hat\Sigma^{(2, t/\tilde t)}_{\phi_1\phi_2}(0) = 0\,, \\
\hat\Sigma^{(2, t/\tilde t)}_{\phi_2}(0) &=
  \frac{G_F\sqrt 2}{\pi^2}\frac{\overline{\alpha}_s(m_t^2)}{\pi}
  \frac{\mtbar^4}{\sinQb} \Biggl[
    3\ln^2\frac{\mtbar^2}{M_S^2}
    - 6\ln\frac{\mtbar^2}{M_S^2}
    - 6\frac{M_t^{LR}}{M_S}
  \nonumber \\
&\hspace*{11em}
    - 3\frac{(M_t^{LR})^2}{M_S^2}\ln\frac{\mtbar^2}{M_S^2}
    + \frac 34\frac{(M_t^{LR})^4}{M_S^4}
  \Biggr]
  \nonumber \\
&\quad
  - \frac{9 G_F^2}{16\pi^4}\frac{\mtbar^6}{\sinQb} \left[
    \tilde X_t\ln\frac{m_{\tilde t_1} m_{\tilde t_2}}{\mtbar^2} +
    \ln^2\frac{m_{\tilde t_1} m_{\tilde t_2}}{\mtbar^2}\right],
\end{align}
where
\begin{align}
\tilde X_t &=
  \left(\frac{m_{\tilde t_2}^2 - m_{\tilde t_1}^2}{\mtbar^2}
     (U^t_{11} U^t_{12})^2\right)^2
  \left(2 - \frac{m_{\tilde t_2}^2 + m_{\tilde t_1}^2}
                 {m_{\tilde t_2}^2 - m_{\tilde t_1}^2}
     \ln\frac{m_{\tilde t_2}^2}{m_{\tilde t_1}^2}\right)
  + 2\frac{m_{\tilde t_2}^2 - m_{\tilde t_1}^2}{\mtbar^2}
    (U^t_{11} U^t_{12})^2
    \ln\frac{m_{\tilde t_2}^2}{m_{\tilde t_1}^2}\,,
\end{align}
and $U^t$ is the stop mixing matrix defined in Eq.\
\eqref{eq:sfmix}. We use the $\MSbar$ top mass
\begin{equation}
\label{eq:mtbar}
\mtbar = \mtbar(m_t)
\approx \frac{m_t}{1 + \frac 4{3\pi}\overline{\alpha}_s(m_t^2)}
\end{equation}
instead of the pole mass to include also the leading
$t/\tilde t$-contributions beyond $\O(\alpha\alpha_s)$.

\item
The one-loop leading-log contributions from all other sectors:
\begin{align}
\hat\Sigma^{(1, \text{rest})}_{\phi_1} &=
  -\frac{G_F M_Z^4}{12\sqrt 2\pi^2} \biggl[
    \left(\frac{12 N_c m_b^4}{M_Z^4\cos^4\!\beta}
      - \frac{6 N_c m_b^2}{M_Z^2\cosQb}
      + P_b + P_f + P_g + P_{2H}\right)\ln\frac{\MSUSY^2}{M_Z^2}
  \nonumber \\
&\hspace*{6em}
    + \theta(M_A - M_Z) (P_{1H} - P_{2H}) \ln\frac{M_A^2}{M_Z^2}
  \biggr] \cosQb
  \nonumber \\
&\quad
  - \frac{G_F N_c m_b^2}{4\sqrt 2\pi^2\MSUSY^2} \biggl[
    \frac{4 m_b^2 A_b M^{LR}_b}{\cosQb}
       \left(1 - \frac{A_b M^{LR}_b}{12\MSUSY^2}\right)
    - M_Z^2 A_b \left(M^{LR}_b + \frac 13 A_b\right)
  \biggr], \\
\hat\Sigma^{(1, \text{rest})}_{\phi_1\phi_2} &=
  -\frac{G_F M_Z^4}{12\sqrt 2\pi^2} \biggl[
    \left(\frac{3 N_c m_b^2}{M_Z^2\cosQb} - P_b - P_f - P_g' -
       P_{2H}'\right)\ln\frac{\MSUSY^2}{M_Z^2}
  \nonumber \\
&\hspace*{6em}
    + \theta(M_A - M_Z) (P_{1H} + P'_{2H}) \ln\frac{M_A^2}{M_Z^2}
  \biggr]\sinb\cosb
  \nonumber \\
&\quad
  + \frac{G_F N_c m_b^2}{8\sqrt 2\pi^2\MSUSY^2} \biggl[
    \frac{4 m_b^2\mu M^{LR}_b}{\cosQb}
      \left(1 - \frac{A_b M^{LR}_b}{6\MSUSY^2}\right) 
  \nonumber \\
&\hspace*{6em}
    - M_Z^2\tanb
      \left(M^{LR}_b (A_b + \mu\cotb) + \frac 13 (\mu^2 + A_b^2)\right)
  \biggr]\,, \\
\hat\Sigma^{(1, \text{rest})}_{\phi_2} &=
  -\frac{G_F M_Z^4}{12\sqrt 2\pi^2} \biggl[
    (P_b + P_f + P_g + P_{2H}) \ln\frac{\MSUSY^2}{M_Z^2}
  \nonumber \\
&\hspace*{6em}
    + \theta(M_A - M_Z) (P_{1H} - P_{2H}) \ln\frac{M_A^2}{M_Z^2}
  \biggr] \sinQb
  \nonumber \\
&\quad
  + \frac{G_F N_c m_b^2}{4\sqrt 2\pi^2\MSUSY^2} \left[
    \frac{m_b^2\mu^2 (M^{LR}_b)^2}{3\MSUSY^2\cosQb}
    + M_Z^2\mu\tanb\left(M^{LR}_b + \frac 13\mu\tanb\right)
  \right]
\end{align}
with
\begin{equation}
\begin{aligned}
M_b^{LR} &= A_u - \mu^*\cotb\,, \\
P_b &= N_c (1 + 4 Q_b\sw^2 + 8 Q_b^2\sw^4)\,, \\
P_f &= N_c (N_g - 1) \bigl(2 - 4\sw^2 + 8 (Q_t^2 + Q_b^2)\sw^4\bigr)
       + N_g (2 - 4\sw^2 + 8\sw^4)\,, \\
P_g &= -44 + 106\sw^2 - 62\sw^4\,, \\
P'_g &= 10 + 34\sw^2 - 26\sw^4\,, \\
P_{1H} &= -9\cos^4 2\beta + (1 - 2\sw^2 + 2\sw^4) \cos^2 2\beta\,, \\
P_{2H} &= -10 + 2\sw^2 - 2\sw^4\,, \\
P'_{2H} &= 8 - 22\sw^2 + 10\sw^4\,, \\[1ex]
Q_t &= \frac 23\,, \qquad
Q_b = -\frac 13\,, \qquad
N_c = 3\,, \qquad
N_g = 3\,.
\end{aligned}
\end{equation}
\end{itemize}

\subsubsection{The case of complex parameters}
\label{sect:complexpara}

The sophisticated calculation of the Higgs masses as in the preceding
section is valid only for real parameters. When complex parameters are 
used, the preprocessor variable {\tt COMPLEX\uscore PARAMETERS} must be
defined (see Table \ref{tab:mssmini}), which makes \mssmini\ use a simple
(one-loop) approximation for the Higgs masses that is valid also for
complex parameters.

For the Higgs masses the dominant one-loop contributions are the $m_t^4$
terms \cite{Da95}. Neglecting the splitting of the $\tilde t$ masses, the
$\phi_2$ self-energy of Eq.\ \eqref{eq:sigphi2} becomes
\begin{equation}
\hat\Sigma^{(1, t/\tilde t)}_{\phi_2}(0) =
  \frac{3 G_F}{2\sqrt 2\pi^2} \frac{\mtbar^4}{\sinQb}
  \ln\left( 1
    + \frac{M_{\tilde Q}^2\, M_{\tilde U}^2}{\mtbar^4}
    + \frac{M_{\tilde Q}^2 + M_{\tilde U}^2}{\mtbar^2}
  \right)
\end{equation}
with the running top mass $\mtbar$ as defined in Eq.\ \eqref{eq:mtbar}.
All other radiative corrections in the Higgs mass matrix are set to zero.

\subsubsection{The charged Higgs bosons}

There are only small radiative corrections for the charged Higgs masses.
We use the following relation, valid for $M_A\sim\O(M_Z)$ \cite{HaHH97},
up to $M_A$-values of 250 GeV:
\begin{align}
M_{H^\pm}^2 &= M_A^2 + M_W^2
  + \frac{\alpha M_W^2}{12\pi\sw^2} \left[
      N_c (N_g - 1) + N_g - 9 + 15\frac{\sw^2}{\cw^2}
  \right] \ln\frac{\MSUSY^2}{M_W^2}
  \nonumber \\
&\qquad
  + \frac{N_c\alpha}{8\pi\sw^2 M_W^2} \left[
      \frac{2\mtbar^2 m_b^2}{\sinQb\cosQb}
      - M_W^2 \left(\frac{\mtbar^2}{\sinQb} + \frac{m_b^2}{\cosQb}\right)
      + \frac 23 M_W^4
  \right] \ln\frac{\MSUSY^2}{\mtbar^2}\,,
\end{align}
with $N_c = N_g = 3$ and the running top mass $\mtbar$ as defined in Eq.\
\eqref{eq:mtbar}. Above 250 GeV, this formula reduces to $M_{H^\pm}^2 =
M_A^2 + M_W^2$.


\subsection{The chargino sector}

The charged gauginos and higgsinos mix via the mass matrix
\cite{HaK85, GuH86}
\begin{equation}
X = \begin{pmatrix}
M_2 & \sqrt2 M_W\sinb \\
\sqrt 2 M_W\cosb & \mu
\end{pmatrix}.
\end{equation}
$X$ is diagonalized with two unitary matrices $U$ and $V$ according to
\begin{equation}
U^* X V^\dagger = \diag(m_{\tilde\chi_1}, m_{\tilde\chi_2})\,,
\end{equation}
which yields the chargino masses
\begin{equation}
m^2_{\tilde\chi_{1,2}} = \frac{M_2^2 + |\mu|^2 + 2 M_W^2}{2} 
  \mp\sqrt{\frac{(M_2^2 + |\mu|^2 + 2 M_W^2)^2}{4} 
       - |M_W^2\sin 2\beta - \mu M_2|^2}\,.
\end{equation}
For the numerical diagonalization, the LAPACK subroutine ZGESVD
\cite{lapack} is used.


\subsection{The neutralino sector}

The neutral gauginos and higgsinos mix via the mass matrix
\cite{HaK85, GuH86}
\begin{equation}
Y = \begin{pmatrix}
M_1          &  0           & -M_Z\sw\cosb &  M_Z\sw\sinb \\ 
0            &  M_2         &  M_Z\cw\cosb & -M_Z\cw\sinb \\
-M_Z\sw\cosb &  M_Z\cw\cosb &  0           & -\mu         \\ 
M_Z\sw\sinb  & -M_Z\cw\sinb & -\mu         &  0
\end{pmatrix}.
\end{equation}
$Y$ is diagonalized with a unitary matrix $N$ according to
\begin{equation}
N^* Y N^\dagger =
\diag(m_{\tilde\chi^0_1}, m_{\tilde\chi^0_2},
      m_{\tilde\chi^0_3}, m_{\tilde\chi^0_4})\,,
\end{equation}
which yields the four neutralino mass eigenstates. For the numerical
diagonalization, the LAPACK subroutine ZGESVD \cite{lapack} is used.
Note that we work with a general complex matrix $N$ and also check that
the computed mass values are nonnegative.


\subsection{The sfermion sector}

The sfermion gauge eigenstates are connected via the mass matrix
\cite{HaK85, GuH86}
\begin{equation}
Z = \begin{pmatrix}
M^{LL}_f + m_f^2 & m_f (M^{LR}_f)^*  \\
m_f M^{LR}_f     & M^{RR}_f + m_f^2
\end{pmatrix}
\end{equation}
where
\begin{align}
M^{LL}_f &= M_Z^2 (I_3^f - Q_f\sw^2) \cos 2\beta + 
\begin{cases}
M^2_{\tilde Q} & \text{for left-handed squarks}, \\
M^2_{\tilde L} & \text{for left-handed sleptons},
\end{cases} \\
M^{LR}_f &= A_f - \mu^*
\begin{cases}
\cotb & \text{for $u$-type sfermions ($I_3^f = +1/2$)}, \\
\tanb & \text{for $d$-type sfermions ($I_3^f = -1/2$)},
\end{cases} \\
M^{RR}_f &= M_Z^2 Q_f\sw^2 \cos 2\beta + 
\begin{cases}
M^2_{\tilde U} & \text{for right-handed, $u$-type squarks}, \\
M^2_{\tilde D} & \text{for right-handed, $d$-type squarks}, \\
M^2_{\tilde E} & \text{for right-handed sleptons}.
\end{cases}
\end{align}
The mass eigenstates are obtained by diagonalizing $Z$ with a unitary
matrix $U^f$, viz.
\begin{equation}
\label{eq:sfmix}
U^f Z U^{f\dagger} =
  \diag(m^2_{\tilde f_1}, m^2_{\tilde f_2})\,,
\end{equation}
which results in the sfermion masses
\begin{equation}
\label{eq:msfermion}
m_{\tilde f_{1,2}}^2 = m_f^2 + \frac 12\left(
  M^{LL}_f + M^{RR}_f
  \mp\sqrt{\bigl(M^{LL}_f - M^{RR}_f\bigr)^2
    + 4 m_f^2 \bigl|M^{LR}_f\bigr|^2}\right).
\end{equation}
For the numerical diagonalization the LAPACK subroutine ZHEEV
\cite{lapack} is used.

Sfermion mixing can be switched off by defining the preprocessor variable
{\tt NO\uscore SQUARK\uscore MIXING} (see Table \ref{tab:mssmini}). This
causes the breaking parameters $A_u$ and $A_d$ to be set to the values
\begin{equation}
\label{eq:nosqmix}
A_u = \mu^*\cotb\,,\qquad A_d = \mu^*\tanb
\end{equation}
so that the off-diagonal terms in the mass matrix $Z$ vanish.

\subsubsection{Supersymmetric contributions to $\Delta\rho$}

In addition to the experimental bounds on the squark masses, \mssmini\
also checks whether the MSSM contributions to the $\rho$-parameter are
consistent with the current exclusion limits (see Eq.\ \eqref{eq:bounds}).
This bound becomes relevant mainly when parameters are chosen to achieve a
large mass splitting between $\tilde t_1$ and $\tilde t_2$ or $\tilde b_1$
and $\tilde b_2$.

We have implemented the calculation of the MSSM contributions to
$\Delta\rho$ according to \cite{DjGHHJW97}. The sfermion contributions are
significant only when the masses of the isospin partners are very
different, and since this is governed by the quark masses (see Eq.\
\eqref{eq:msfermion}), only corrections from $\tilde t/\tilde b$-loops
have been included. Contributions from gluino exchange, which are very
lengthy and vanish for large $m_{\tilde g}$, have similarly been
neglected.

These contributions to $\Delta\rho$ are given by
\begin{align}
\Delta\rho_{\tilde t,\tilde b} &=
    U^t_{11} U^t_{22} U^{t*}_{12} U^{t*}_{21}
      F(m_{\tilde t_1}^2, m_{\tilde t_2}^2)
  + U^b_{11} U^b_{22} U^{b*}_{12} U^{b*}_{21}
      F(m_{\tilde b_1}^2, m_{\tilde b_2}^2)
    \nonumber \\
&\qquad
  + |U^t_{11}|^2 |U^b_{11}|^2
      F(m_{\tilde t_1}^2, m_{\tilde b_1}^2)
  + |U^t_{11}|^2 |U^b_{21}|^2
      F(m_{\tilde t_1}^2, m_{\tilde b_2}^2)
    \nonumber \\
&\qquad
  + |U^t_{21}|^2 |U^b_{11}|^2
      F(m_{\tilde t_2}^2, m_{\tilde b_1}^2)
  + |U^t_{21}|^2 |U^b_{21}|^2
      F(m_{\tilde t_2}^2, m_{\tilde b_2}^2)\,,
\end{align}
where the $U^f$ are the sfermion mixing matrices defined in Eq.\
\eqref{eq:sfmix} and the function
\begin{equation}
F(x, y) = \frac{3 G_F}{8\sqrt 2\pi^2} F_1(x, y)
  + \frac{G_F}{4\sqrt 2\pi^2}
    \frac{\overline{\alpha}_s(m_t^2)}{\pi} F_2(x, y)
\end{equation}
includes the one- and two-loop contributions
\begin{align}
F_1(x, y) &= x + y - \frac{2 x y}{x - y}\ln\frac xy\,, \\
F_2(x, y) &= x + y
- \frac{2 x y}{x - y}\left(2 + \frac xy\ln\frac xy\right)\ln\frac xy
+ \frac{x^2 (x + y)}{(x - y)^2} \ln^2\frac xy
- 2 (x - y) \Li_2\!\left(1 - \frac xy\right).
\end{align}
The two-loop contribution is of the order of 10 to 15\% of the one-loop
result.


\section{Tests}
\label{tests}

The model files {\tt MSSM.mod} and {\tt MSSMQCD.mod} have been checked
against results known from the literature for a variety of scattering
processes. In cases where the Fortran programs of the original authors
were available, we found perfect agreement for all differential
cross-sections. In the other cases we could reproduce qualitatively the
figures. The list of processes we checked together with the agreement we
achieved is given in Table \ref{tab:checks}.

\begin{table}
\begin{center}
\renewcommand{\arraystretch}{1.5}
\begin{tabular}{|c|c|c|c|} \hline
process & agreement & reference & scope \\ \hline
$q\bar q\to t\bar t$ & 11 digits & \cite{Be00} &
	SUSY-QCD only \\
$q\bar q\to t\bar t$ & 10 digits & \cite{Wa01} &
	full MSSM \\
$gg\to t\bar t$ & qualitative & \cite{ZePWGLY99} &
	SUSY-QCD only \\
$gg\to t\bar t$ & 10 digits & \cite{Wa01} &
	full MSSM \\
$gg\to H^+H^-$ & 10 digits & \cite{BrH00} &
	full MSSM \\
$gg\to W^-H^+$ & qualitative & \cite{BrHK01} &
	full MSSM \\
$e^+e^-\to t\bar t$ & 11 digits & \cite{HoS99} &
	full MSSM \\
$e^+e^-\to W^+W^-$ & qualitative & \cite{AlHKSU00} &
	sfermion contributions \\
$e^+e^-\to H^+H^-$ & 7 digits & \cite{GuHK01} &
	full MSSM \\
dipole moments & 11 digits & \cite{HoIRSS99} &
	full MSSM \\ \hline
\end{tabular}
\end{center}
\caption{\label{tab:checks}Checks performed with \FA\ and \FC\ to test the
implementation of the MSSM.}
\end{table}


\section{Availability, Requirements}
\label{sect:download}

The \FA\ package including the {\tt MSSM.mod} and {\tt MSSMQCD.mod} model
files can be downloaded from {\tt http://www.feynarts.de}. The program
itself requires \mma\ 3 or above. For the topology editor, which is not
necessarily invoked, a Java VM and the J/Link package are needed, both of
which can be obtained free of charge (see the instructions on the web
site).

The \FC\ package is available from {\tt http://www.feynarts.de/formcalc}
and includes the initialization file \mssmini. It runs on Unix-like
platforms and requires \mma\ 3, \FO\ 3, and the GNU C compiler (gcc). To
link the Fortran code generated by \FC, one needs in addition the \LT\
library ({\tt http://www.feynarts.de/looptools}) and the CERNlib
({\tt http://wwwinfo.cern.ch/asd/index.html}).

\FA\ and \FC\ each include a comprehensive manual which explains
installation and usage. Both are open-source programs and stand under the
GNU library general public license.

\section*{Acknowledgements}

We thank A.~Kraft for his Fortran code which provided the basis for
{\tt MSSM.mod}, S.~Berge, O.~Brein, and D.~Wackeroth for helping us
cross-check our programs, W.~Hollik and S.~Heinemeyer for proofreading the
manuscript and valuable suggestions on the implementation of the Higgs
masses, and G.~Jahn for some checks of the four-sfermion couplings.

T.H. has been supported by the Deutsche Forschungsgemeinschaft
(Forschergruppe ``Quantenfeldtheorie, Computeralgebra und Monte-Carlo
Simulation'') under contract number Ku 502/8--1.


\newcommand{\cpc}[3]{{\sl Comp. Phys. Commun.} {\bf #1} (#2) #3}
\newcommand{\epj}[3]{{\sl Eur. Phys. J.} {\bf #1} (#2) #3}
\newcommand{\fp}[3]{{\sl Fortschr. Phys.} {\bf #1} (#2) #3}
\newcommand{\np}[3]{{\sl Nucl. Phys.} {\bf #1} (#2) #3}
\newcommand{\npps}[3]{{\sl Nucl. Phys. Proc. Suppl.} {\bf #1} (#2) #3}
\newcommand{\pl}[3]{{\sl Phys. Lett.} {\bf #1} (#2) #3}
\newcommand{\prep}[3]{{\sl Phys. Rep.} {\bf #1} (#2) #3}
\newcommand{\pr}[3]{{\sl Phys. Rev.} {\bf #1} (#2) #3}
\newcommand{\prl}[3]{{\sl Phys. Rev. Lett.} {\bf #1} (#2) #3}
\newcommand{\zp}[3]{{\sl Z. Phys.} {\bf #1} (#2) #3}
\newcommand{\nim}[3]{{\sl Nucl. Instr. Meth.} {\bf #1} (#2) #3}
\newcommand{\jcp}[3]{{\sl J. Comput. Phys.} {\bf #1} (#2) #3}
\newcommand{\ptps}[3]{{\sl Prog. Theor. Phys. Suppl.} {\bf #1} (#2) #3}

\end{document}